\documentclass[useAMS,usenatbib,A4Paper]{mn2e}

\usepackage{longtable}
\usepackage{graphicx}

\title[Disc Heating]{Disc Heating: Comparing the Milky Way
 with Cosmological Simulations}

\author[E.L. House et~al.]{E.L. House$^{1}$, C.B. Brook$^{1}$,
        B.K. Gibson$^{1}$, P.S\'{a}nchez-Bl\'{a}zquez$^{1,2}$, S.~Courty$^{1,3}$,
        \newauthor C.G. Few$^{1}$, F. Governato$^{4}$, D. Kawata$^{5}$, 
        R. Ro\v{s}kar$^{4,7}$, M. Steinmetz$^{6}$, 
        \newauthor G.S. Stinson$^{1}$ and R. Teyssier$^{7,8}$ \\
        $^1$University of Central Lancashire, Jeremiah Horrocks 
            Institute, Preston, PR1~2HE, UK \\
        $^2$Grupo de Astrof\'{i}sica,
	    Departamento de Fisica Teorica, 
	    Universidad Aut\'onoma de Madrid, Cantoblanco,
            E-28049, Spain\\
	$^3$Centre de Recherche Astrophysique de Lyon, UMR 5574,
	    9 Avenue Charles Andr\'e, F69561 Saint Genis Laval, France\\
        $^4$Astronomy Department, University of Washington, 
            Box 351580, Seattle, WA 98195-1580, USA \\
        $^5$Mullard Space Science Laboratory, University College London, 
            Holmbury St. Mary, RH1~6NT, UK \\
        $^6$Astrophysikalisches Institut Potsdam, 
            An der Sternwarte 16, 14482 Potsdam, Germany \\
        $^7$Institute for Theoretical Physics, University of Z\"{u}rich, 
            CH-8057, Z\"{u}rich, Switzerland \\
        $^8$UMR AIM, CEA Saclay, 
            91191 Gif-sur-Yvette, France}

\begin{document}

\date{\today}

\maketitle
\begin{abstract}
We present the analysis of a suite of simulations run with different 
particle-and grid-based cosmological hydrodynamical codes and compare 
them with observational data of the Milky Way. This is the 
first study to make comparisons of properties of galaxies simulated  with particle 
and grid-based codes. Our analysis indicates that there is broad agreement 
between these different modelling techniques. We study the velocity 
dispersion -- age relation for disc stars at $z=0$ and find that four 
of the simulations are more consistent with observations by \cite{holmberg08}
 in which the stellar disc appears to undergo 
continual/secular heating. Two other simulations are in better agreement 
with the \cite{quillen01} observations that suggest a 
``saturation'' in the heating profile for young stars in the disc. 
None of the simulations have thin discs as old as that of the Milky Way.
We also analyse the kinematics of disc stars at the time of their birth for 
different epochs in the galaxies' evolution and find that in some 
simulations old stars are born cold within the disc and are subsequently 
heated, while other simulations possess old stellar populations which 
are born relatively hot. The models which are in better agreement with 
observations of the Milky Way's stellar disc undergo significantly lower 
minor-merger/assembly activity after the last major merger -- i.e. once 
the disc has formed.  All of the simulations are significantly 
``hotter'' than the Milky Way disc; on top of the effects of mergers, 
we find a ``floor'' in the dispersion that is related to the underlying treatment 
of the heating and cooling of the interstellar medium, and the low density 
threshold which such codes use for star formation. This finding has 
important implications for all studies of disc heating that use 
hydrodynamical codes.  \end{abstract}

\begin{keywords}
galaxies: formation---galaxies: evolution---Galaxy: thick and thin disc---
methods: N-body simulations
\end{keywords}

\section{Introduction}

One of the major outstanding ``grand challenges'' facing astrophysics 
for the coming decade is the unravelling of the underlying physics 
governing the formation and evolution of disc galaxies such as our own 
Milky Way. A principal difficulty resides in trying to accommodate the 
early collapse and violent merging history intrinsic to the canonical 
framework of ``hierarchical assembly'' of galactic structure with the 
apparent stability of what should be fairly fragile thin galactic discs.

High Performance Computing (HPC) simulations of gravitational {\it 
N}-body and hydrodynamical physics have become a primary tool with which 
to model galaxy formation in a cosmological context (e.g.\cite{katz92}; 
\cite{summers93}; \cite{navarro94}; \cite{steinmetz94};
\cite{sommer03}; \cite{abadi03a}; \cite{robertson04}; 
\cite{bailin05}; \cite{okamoto05}; \cite{g07}; \cite{gibson09}; 
\cite{sanchez09}; \cite{agertz10}). These simulations 
model the formation and evolution of disc galaxies within a Universe 
dominated by a Cold Dark Matter (CDM) component and a cosmological 
constant ($\Lambda$). While powerful, the techniques employed are not 
without their problems; for example, the loss of angular momentum in the 
luminous component of disc galaxies is one of the major problems in most 
of the aforementioned cosmological simulations. In these simulations, 
gas cools efficiently via radiative processes, causing baryons to 
collapse rapidly during the earliest phases of the hierarchical clustering 
process. The luminous component ends up transferring angular momentum to 
the dark matter halo making the luminous component deficient in 
angular momentum. This is often referred to as the ``angular momentum problem''
(\cite{navarro91}; \cite{steinmetznavarro02}).
As a result, these simulations typically produce galaxies with an 
overly-dominant spheroid component and an overly small disc (\cite{abadi03a}; 
\cite{scannapieco09}), 
in disagreement with observations of disc galaxies (\cite{brook04a}).

Another challenge facing disc galaxy formation in the $\Lambda$CDM 
scenario is the old age of the Milky Way's thin disc. This seems at odds 
with the heating that one expects from merging and accretion events 
within a $\Lambda$CDM paradigm. Indeed, several studies of isolated 
discs being bombarded by satellites have shown that one would expect 
that the disc would be destroyed, or at least severely heated, by 
accretion events (\cite{quinn93}; \cite{kazantzidis08}; \cite{kazantzidis09}; 
\cite{read08}). Two recent studies have included gas in the main disc, 
with one (\cite{moster10}) finding a significant decrease in heating, 
by $25\%-40\%$ for gas fractions of $20\%$ and $40\%$ respectively, with 
the other (\cite{purcell10}) finding that the effects of gas are 
somewhat less dramatic. What is clear is that all studies which use 
contrived initial conditions which are bombarded with satellites are 
necessarily restricted in their application, both for the disc and 
satellites. For example, how best to assign an appropriate velocity 
dispersion, mass, and scalelength of the Milky Way disc at redshift of 
two, say? Were the stars already kinematically ``hot'' in this early 
disc, or have they been heated subsequently? Idealised studies with 
pre-formed discs can be powerful, but they do not address directly the 
issues pertaining to disc formation and how this relates to merger 
events as they occur within a hierarchical cosmology. Suffice to say 
that the existence of thin discs remains a challenge for $\Lambda$CDM 
cosmology. \cite{stewart09} argue that gas rich-mergers can explain 
the number of low mass galaxies on the blue sequence and mass-morphology 
relation, but their analysis is not able to address the issue of the 
thinness of the discs which survive mergers. In fact, it has been shown 
(\cite{brook04}; \cite{springel05}; \cite{robertson08}; 
\cite{g09}) that gas rich mergers in simulations result in 
hot thick discs, with thin discs forming in the subsequent quiescent period.

Observations of the kinematics of disc stars of our Galaxy have been 
carried out throughout the years in order to understand the mechanisms 
governing the formation of the disc. These studies include \cite{nordstrom04} 
and the follow up study by \cite{holmberg07}, \cite{soubiran05}, \cite{soubiran08}, 
 \cite{quillen01}, and \cite{dehnen98}. These studies however, 
have provided different pictures of the relationship between the ages of 
disc stars and their velocity dispersions (the age-dispersion relation). 
\cite{quillen01}, using the data of \cite{edvardsson93}, found 
that vertical disc heating for the Milky Way saturates at 
$\sigma_{w}\sim20$ kms$^{-1}$, with the value of dispersion virtually 
constant for stars of ages between $\sim$2 and $\sim$9~Gyr. A discrete 
jump is apparent for stars with ages $>$9~Gyr which is generally 
interpreted to be the signature of the thick disc. Thus, this study 
supports the notion of a thick disc as a separate component to the thin 
disc, and suggests different formation scenarios for each component. By 
contrast, \cite{nordstrom04} and the follow up study of \cite{holmberg07} 
advocate a picture in which the disc has undergone continual 
heating over the past $\sim$10~Gyr. It is not clear from these later 
studies whether a thick disc should be considered as a separate 
component: firstly, the selection is biased toward thin disc stars, and 
secondly, if the thick disc is a separate component, it is possible that 
their result is driven by increasing contamination of their sample by 
thick disc stars as older and older stars are examined (\cite{navarro10}). 
One of the main 
points of contention in such studies, and a possible explanation for the 
different findings, is the difficulty in determining the ages of stars 
(\cite{anguiano09}; \cite{aumer09}). Further, \cite{seabroke07} 
showed that a power law fit as suggested in the 
Geneva-Copenhagen studies is statistically similar to disc heating 
models which saturate after $\sim$4.5~Gyr, and is consistent with a 
minimal increase of $\sigma_{w}$ for old stars. We will compare our 
simulations to both these data sets.

The degree of heating of thin disc stars is certainly complicated by an 
old, hot ``thick disc'' of stars. Since the discovery of the thick disc 
component of the Milky Way by \cite{gilmore83}, several analyses 
of ages, abundances, and kinematics have indicated that the thick and 
thin discs are two distinct components (e.g \cite{reid93}; 
\cite{nissen95}; \cite{chiba00}; \cite{bensby07}; \cite{beers09} 
although see \cite{ivezic08} for an opposing view), where a thin disc whose 
stars have formed continuously over $\sim9$~Gyr is superimposed on an 
old thick disc. The thick disc has also been shown to be a common 
component in most, and possibly all, observed spiral galaxies (\cite{yoachim08a}), 
at least in the sense that the light distributions 
are better fit by two functions rather than one.  Recent observations 
carried out measuring the stellar population of the Milky Way's thick 
disc give scaleheights that range between 500 - 1100~pc (\cite{juric08};  
\cite{carollo10}) compared to the thin disc that has a measured 
scaleheight that ranges between 200 - 400~pc (\cite{juric08}; 
\cite{carollo10}). The rotational lag of the thick disc of the Milky 
Way ranges from 20 kms$^{-1}$ (\cite{chiba00}) to 50 kms$^{-1}$ 
(\cite{soubiran08}). The velocity ellipsoid (i.e., the velocity 
dispersions in the $u$,$v$,$w$ reference frame) of the thick disc has 
been quoted as being from ($\sigma_{u},\sigma_{v},\sigma_{w})=(46 
\pm4,50 \pm4,35 \pm3)$ kms$^{-1}$, as measured by \cite{chiba00}, 
to $(\sigma_{u},\sigma_{v},\sigma_{w})=(63 \pm6,39 \pm4,39 \pm4$)
kms$^{-1}$, as measured by \cite{soubiran08}. The thin disc 
dominates the thick disc in the local region by a factor of 10:1 in 
terms of stellar mass, but difficulty in determining scalelengths of the 
two means that comparing their total masses is highly uncertain, with 
estimates ranging from mass ratios of 10:1 to 3:1 \cite{juric08}.
  
For this study we focus on the kinematics and heating of all stars 
within the disc region of our simulations, without any {\it a priori} 
distinction between the thin and thick discs.  We will explore and 
discuss occasions where two components arise, in the hope of shedding 
light on the various scenarios which have been postulated to explain the 
formation of the thick and thin discs. One model suggests that a 
previously existing thin disc was kinematically heated to give rise to 
the thick disc. This vertical heating of the thin disc stars could be 
rapid, due to mergers (\cite{quinn93}), or more gradual, as a result 
of secular processes such as giant molecular clouds, spiral arms or the 
presence of a bar providing the necessary kinematic ``kick'' to the 
stars (\cite{larson76}). An alternative model suggests that the thick disc 
formed during the violent relaxation of the galactic potential prior to 
the formation of the thin disc, where star formation was triggered by 
accretion of gas during major merger events at an early epoch (\cite{brook04}) . 
This raises the question of whether the age-dispersion 
relation is at least in part due to the earlier discs being, in general, 
hotter than the later discs -- i.e. the old stars were born hotter than 
the younger stars, possibly related to earlier mergers in $\Lambda$CDM 
having higher mass ratios between the mass of the central galaxy to the 
accreted satellite in general than later mergers (\cite{brook05}). 
Recent observations suggesting that high-redshift discs are relatively 
thick (\cite{dalcanton02}; \cite{vanstarkenburg08}; 
\cite{lemoine10}) possibly provide support for this 
scenario. A further alternative model suggest that the thick disc stars 
are actually accreted from satellite galaxies during the hierarchical 
assembly process (\cite{abadi03a}). Recently, models of discs which are 
entirely isolated from the satellite bombardment that is predicted in 
$\Lambda$CDM have shown that migration of stars can naturally lead to 
combinations of age, metallicity and dispersions which are consistent 
with observations of thin and thick disc populations (\cite{loebman10} 
as well as an analytic model by \cite{schonrich09}). 
These models beg the question of whether heating is required at all, 
yet to be fair, they have not been integrated within a fully cosmological 
paradigm. Finally, \cite{kroupa02} suggested that massive star clusters 
formed at high redshift dissolve at later times to form the thick disc, a 
theory given support by the clumpy nature of high redshift discs 
(\cite{elmegreen07}; \cite{elmegreenelmegreen06}). 

We aim to provide further insight into the causes of the age-velocity 
dispersion relation of all disc stars (thick and thin) by using a suite 
of HPC simulations of disc galaxies, each run with a different 
hydrodynamical code, different initial conditions, different resolution, 
and different assembly histories, to sample a wide range of disc galaxy 
formation pathways. In particular, we determine whether disc stars are 
born hot or cold in early times, and the degree in which they 
subsequently heat vertically. Attention is given to any connection 
between heating rates and accretion histories. The issue of the effects 
of numerical heating is important in all studies of disc heating. We 
show that numerical heating is not causing the measured disc heating. We 
highlight the role of the implementation of star formation recipes, and 
the modelling of the interstellar medium in which stars are formed, as 
determinants of a dispersion ``floor'' for the simulations. 

In \S~2, we briefly describe the main characteristics of each of the 
different codes used to produce our compilation of simulations. In \S~3, 
we present our main results which include the dispersion of all stars as 
a function of time for the final galaxies, comparing the simulations 
with observations of the Milky Way, studying the kinematics of young 
stars at different epochs, as well as following the heating of stars 
born at early epochs; we then relate the heating with merger processes 
within the simulations. In \S~4, we examine isolated simulations, and 
show that heating does not occur in the absence of a cosmological 
environment, ruling out numerical effects as the primary agent driving 
our results. We present our conclusions in \S~5.

\section{The Simulations}

\begin{table*}
\caption{Summary of the properties for the simulation suite}
\label{Tabel 1}
\begin{tabular}{|c|c|c|c|c|c|c|c|c|c|}
\hline
Simulation Name & Code & $M_{vir}$        & $\Omega_0$ & $h_0$ & $\Omega_b$ & $t_{LMM}$ & $\epsilon$& Gas Resolution & Feedback\\
                &      & ($~M_\odot$)     &            &       &            &    Gyr   & (kpc)     & ($~M_\odot$)   &          \\
\hline
  S09 & {\tt RAMSES} & 7.6$\times$10$^{11}$ & 0.3 & 0.7 & 0.045  & 10.99 & 0.4 & $1.0\times10^{6}$ & kinetic\\
  G07(MW1) & {\tt GASOLINE} & 1.1$\times$10$^{12}$ & 0.3 & 0.7 & 0.039  & 10.89 & 0.6 & $8.0\times10^{5}$ & adiabatic\\
  B09(h277) & {\tt GASOLINE} & 7.1$\times$10$^{11}$ & 0.24 & 0.77 & 0.045  & 10.41 & 0.35 & $1.6\times10^{4}$ & adiabatic\\
  B05(SGAL1) & {\tt GCD+} & 5.0$\times$10$^{11}$ & 1 & 0.5 & 0.1  & 9.74 & 0.6 & $1.0\times10^{6}$ & adiabatic\\
  A03 & {\tt GRAPESPH} & 9.4$\times$10$^{11}$ & 0.3 & 0.65 & 0.045  & 8.31 & 0.5 & $2.0\times10^{6}$ & kinetic\\
  H09 & {\tt RAMSES} & 7.6$\times$10$^{11}$ & 0.3 & 0.7 & 0.045  & 10.99 & 0.2 & $1.3\times10^{5}$ & adiabatic\\
  R08 & {\tt GASOLINE} & 1.0$\times$10$^{12}$ & 0.3 & 0.7 & 0.039  & N/A & 0.05 & $0.2\times10^{5}$ & adiabatic\\
\hline
\end{tabular}
\label{simproperties}
\end{table*}

We analyse seven cosmological disc simulations run with different {\it 
N}-body hydrodynamical galaxy formation codes. In this section we 
provide a summary of the main details of each code. For full details, 
references are provided.

Two of the simulations we analysed in this study are run with {\tt 
RAMSES} (\cite{teyssier02}), 
which models the gas hydrodynamics using an adaptive mesh 
refinement (AMR) scheme, while the other codes use a smoothed particle 
hydrodynamics (SPH) approach.\footnote{\cite{knebe05} provides an 
excellent primer to the differences between the particle- and 
grid-based approach to solving Poisson's Equation, in an
$N$-body context.}
Using examples drawn from these different 
fundamental approaches should provide greater confidence in the 
robustness of our results.
To the best of our knowledge, our study is the first to 
compare properties of simulated disc galaxies formed 
using these two commonly-adopted methodologies.

All five simulations\footnote{As we mention below, we actually
analyse six simulations, but one of these is simply a higher-resolution 
version of one of the base simulations, and so is not entirely
``independent''. We also include in our analysis a non cosmological 
simulation, see \S~2.7.} have cosmological initial conditions where small 
scale structures merge to form increasingly larger objects in the 
Universe as part of the so-called hierarchical framework. 
They all have a similar 
Milky Way-type mass halo and all the different codes self-consistently 
include the primary physical processes needed to model galaxy formation 
and evolution. These consist of the effects of gravity, hydrodynamic 
pressure and shocks, star formation and feedback, radiative cooling, and 
a photoionising UV background. They all adopt a type of Schmidt law 
for converting gas particles into stars, where the star formation rate 
is proportional to the gas density to some power.

The main difference between the simulations is that they have different 
initial conditions, and hence merger and assembly histories. They also 
adopt different recipes for feedback from Type~II supernovae (SNeII). 
Various methods have been suggested
to incorporate the supernovae feedback into numerical 
simulations: one technique is to artificially turn off radiative 
cooling in the area where the SNeII explosion occurs for 
a timescale long enough to allow the blast wave to expand. 
We call this type of feedback ``adiabatic feedback''. The 
second approach is to directly inject kinetic energy into the 
surrounding gas; we refer to this as ``kinetic feedback''. Two of our 
simulations use kinetic feedback, while the rest use an adiabatic 
approach. Pure ``thermal feedback'' is used in the case of 
Type~Ia supernovae (SNeIa),
where the longer lifetimes of the progenitors (relative to SNeII)
means that the energy is not 
released into the same high density regions from which the
stellar particles formed (and hence, the associated energy is 
not radiated away as efficiently as for the case of SNeII).

Each of the simulations employ star formation recipes which are
similar; stars can form only
from gas above a certain density threshold. Since cosmological simulations 
typically lack resolution below a few hundred parsecs, this sets a 
maximum density that the simulations can resolve on the order 
of 0.1~cm$^{-3}$; all of the simulations here adopted this star
formation threshold.  We shall discuss the impact of this threshold
selection later in the paper.

The main properties of each of our simulations are presented below, and 
summarised in Table~1.
 
\subsection{S09\_YCosm\_AMR\_RAMSES}

We ran a high-resolution fully-cosmological (YCosm) disc simulation to 
redshift zero using the adaptive mesh refinement (AMR)-based 
code {\tt RAMSES} (\cite{teyssier02}). 
The supernova feedback (SN) is modelled by directly 
injecting kinetic energy into the surrounding gas -- i.e., kinetic 
feedback.

The simulation (S09, hereafter) 
was run within a ``concordance'' cosmology framework, with
$\Omega_0=0.3$, $h_0=0.7$, 
$\Omega_b=0.045$, and $\Omega_\Lambda=0.7$.
Preliminary analysis for this simulation was presented in 
\cite{gibson09}, while its optical properties were categorised extensively
by \cite{sanchez09}. The simulated disc had 
its last major merger (LMM, defined as having total mass ratio of 1:3 or higher) 
at a redshift of $z=2.6$, i.e. $t_{LB}=10.99$ Gyr
(where $t_{LB} =$ lookback time), however, 
interactions with smaller satellites still occur at lower redshifts. 
We discuss the LMM later in the paper. 
Its final virial mass is 7.6$\times$10$^{11}$~M$_\odot$ at $z=0$.

\subsection{G07\_MW1\_YCosm\_SPH\_GASOLINE}

This galaxy is the simulation denoted as MW1 in the work of 
\cite{g07}, and is referred to as 
G07(MW1), hereafter. The code used for this fully-cosmological 
(YCosm) simulation is the smoothed particle hydrodynamics (SPH) code {\tt 
GASOLINE} (\cite{wadsley04}). The SN feedback mechanism 
uses an adiabatic feedback approach where cooling was stopped artificially
to allow blast waves from SNe to expand and heat the surrounding interstellar
medium (ISM). In all the simulations run with {\tt GASOLINE} presented here,
40\% of the SNe energy is coupled to the surrounding gas. Such a prescription
results in a decrease in the amount of gas cooling early 
in the galaxy's formation, reducing the loss of angular momentum 
resulting from the merging of dense stellar systems.

The simulation employed was run within a concordance cosmology with
$\Omega_0=0.3$, $h_0=0.7$, $\Omega_b=0.039$, 
and $\Omega_\Lambda=0.7$; the last major merger was at redshift $z$=2.5, i.e.
$t_{LB}=10.89$ Gyr,
with several late minor interactions thereafter.
The final virial mass is 1.1$\times$10$^{12}$~M$_\odot$.

\subsection{B09\_h277\_YCosm\_SPH\_GASOLINE}

This simulation was also run with {\tt GASOLINE} and was previously studied 
in \cite{brooks09}, and is referred to as B09(h277),
hereafter. The simulation was run in a concordance cosmology,
with $\Omega_0=0.24$, $h_0=0.77$, $\Omega_b=0.045$, and
$\Omega_\Lambda=0.76$. The redshift of its LMM is also at $z$=2.5, i.e. 
$t_{LB}=10.41$ Gyr, but unlike
the case for G07(MW1), this simulation experiences no mergers or
accretion events since $z$$\approx$0.7 or $t_{LB} = 5.8$ Gyr. 
The force resolution is also somewhat higher than for G07(MW1).

\subsection{B05\_SGAL1\_NCosm\_SPH\_GCD+}

This simulation was taken from the work of \cite{brook05}, and
is referred to as
B05(SGAL1), hereafter. The SPH code {\tt GCD+} (\cite{kawata03}) 
was employed, although this
particular run was not fully cosmological (NCosm).
Semi-cosmological models, like B05(SGAL1),
consist of an isolated sphere of dark matter and gas instead 
of a large cosmological volume.
Small-scale fluctuations are superimposed on the sphere to allow for 
local collapse and subsequent star formation. Solid-body rotation is 
also applied to the sphere to incorporate the effects of longer 
wavelength fluctuations that a semi-cosmological model does not 
otherwise account for. Feedback from SNeII was assumed to be adiabatic, 
with cooling turned off in the surrounding gas.

The cosmological framework in which B05(SGAL1) was run is quite
different from the simulations discussed thus far; specifically, it 
used $\Omega_0=1$, $h_0=0.5$, $\Omega_b=0.1$, and $\Omega_\Lambda=0$.
While using the currently favoured $\Lambda$CDM framework
would have a significant impact upon simulations of 
large-scale structure formation from Gaussian random noise initial 
conditions, it has been shown that, within the context of single galaxy 
formation models such as B05(SGAL1), the resulting differences are
negligible (\cite{brook05}). In terms of its merger history, 
B05(SGAL1) is not dramatically different from B09(h277), in the sense
of their being little or no merger activity since redshift $z$$\approx$0.5
or $t_{LB}=5.9$ Gyr. 
  
\subsection{A03\_YCosm\_SPH\_GRAPESPH}

This simulation (hereafter referred to as A03) was first
presented in \cite{abadi03a}. It is a 
fully-cosmological (YCosm) Milky Way-like disc galaxy, simulated with the 
{\tt GRAPESPH} code (\cite{steinmetz96}). Feedback is predominantly 
thermal, with 5\% of supernova energy converted into kinetic feedback
and injected into the surrounding gas particles. A flat $\Lambda$CDM
cosmology was assumed, with
$\Omega_0=0.3$, $h_0=0.65$, $\Omega_b=0.045$, and 
$\Omega_\Lambda=0.7$. Its final 
virial mass was 9.4$\times$10$^{11}$~M$_\odot$ and its last major
merger occurred at
$z=1$, i.e. $t_{LB}=8.31$ Gyr,
although a number of minor interactions occur thereafter.

\subsection{H09\_YCosm\_AMR\_RAMSES}
 
This simulation (H09, hereafter) traces the same halo as 
S09\_YCosm\_AMR\_RAMSES described in \S~2.1, but run with a higher spatial 
resolution, and employing a different feedback mechanism for SNeII. Instead of 
using kinetic feedback as in S09, it relies on an adiabatic feedback 
scheme.

\subsection{R08\_NCosm\_SPH\_GASOLINE}

This simulation (R08, hereafter) is taken from \cite{roskar08}.
It is an isolated Milky Way-type disc galaxy 
(1.0$\times$10$^{12}$~M$_\odot$) with solid body rotation added, similar 
to the B05\_SGAl1\_NCosm\_SPH\_GCD+ simulation; where it differs from
the latter is that R08 does not
incorporate small scale density fluctuations. This means that this 
isolated simulation experiences no merger or accretion events. It is evolved 
for 10~Gyr using the same {\tt GASOLINE} code as G07(MW1) and B09(h277), but
at extremely high spatial resolution (softening length of 50~pc).

\section{Results}

For observed solar neighbourhood stars, it is well established that 
there is a relationship between their velocity dispersion and age. We 
refer to the increase of the dispersion with time as {\it disc heating}, 
where the relationship for the solar neighbourhood indicates that the 
older disc stars are kinematically ``hotter'' than its younger
counterparts. Examining the dynamics of 
stars as a function of time therefore contains valuable information 
about the heating processes - driven by some combination of secular 
and satellite merger-driven phenomena.
We focus on the vertical heating ($\sigma_w$, perpendicular to the plane 
of the galaxy) as this out-of-plane heating is 
more susceptible to mergers/interactions. In-plane heating ($\sigma_u$ 
and $\sigma_v$) is more sensitive to spiral wave and bar-driven
heating which we do 
not consider in our study.\footnote{Further, \cite{seabroke07} showed 
that dynamical streams can contaminate the local in-plane velocity 
distributions, which can complicate and compromise the comparison
with simulated in-plane velocity distribution functions which do not 
capture adequately structure on that scale.}

In what follows, we first 
examine the velocity-dispersion age relation, similar to the manner
by which observers study the same relation within the Milky Way, but now
for stars within the simulated discs at
$z$=0. However, since the dispersion of
stars at $z$=0 does not provide direct information as to the 
velocity dispersion of the population \it at birth\rm, we extend our analysis
to study the time evolution of the heating of sub-populations of
disc stars. Specifically, we will attempt to ascertain whether
stars which are kinematically hot \it today \rm were born ``hot'' or 
were born ``cold and heated'' (by whatever means), thereafter.

\subsection{Age-Velocity Dispersion Relation} 

We first examine the velocity dispersion perpendicular to the plane of the 
galaxy ($\sigma_w$) for all stars at $z=0$ within the ``local'' disc, 
which we define as 7$<$$R_{xy}$$<$9~kpc and $|z|<1$~kpc, as a function of 
age (Fig~1).\footnote{In the sense that ``young stars'' have ``small ages'', 
in Fig~1, and ``old stars'' have ``large ages''.} With such a selection
function, all the simulations show little in the way of evidence
for stellar heating for young stars with ages $\sim$1 to $\sim$3~Gyr, 
as well as much higher dispersions for old stars (consistent with
observations of the Milky Way).

We should point out that for the semi-cosmological simulation, 
B05(SGAL1), we used a slightly larger radial cut of 4$<$$R_{xy}$$<$8~kpc 
and $|z|<1$~kpc due to the smaller number of star particles in this particular run.
We found that in the smaller region of 7$<$$R_{xy}$$<$9 kpc, 
there were not enough stars for some age bins to measure an accurate velocity 
dispersion. We were therefore forced to use a larger region for this 
simulation. However, we also tested our results in all the simulations for three different 
radial cuts to ensure that the trends remain the same independent of radius.

The S09 and H09 simulations show little sign of heating for stars younger
than $\sim$3~Gyr; for older stars, several discrete ``jumps'' in
velocity dispersion can be discerned.  Similar trends and 
small discrete jumps are also seen in the G07(MW1) simulation,
in addition to a discontinuity in the dispersion for stars of age
$\sim$8~Gyr.  The latter can be traced to a period of enhanced
merger activity early in the galaxy's evolution, just prior to the
establishment of its stable disc.

Like all the simulations, A03 also shows dispersion trends consistent
with the signature of continual/secular heating at later times. Stars older than
$\sim$9~Gyr, in particular, possess a significantly large velocity dispersion.
These high dispersions are a signature
of the so-called ``angular momentum problem'' mentioned in\S~1, which results in 
the formation of an overly-dominant
spheroid compared to observations. The spheroid 
component is dominated by old stars which are ultimately the 
responsible agents in the production of the high dispersions
seen to the right-hand side of Fig~1.

The absence of significant heating seen in all 
simulations for stellar ages of $\sim$1-3~Gyr extends to somewhat
older stars (up to $\sim$6~Gyr in age), for both B09(h277) and B05(SGAL1).
For older stars, discrete jumps in the dispersion, superimposed upon
a continual heating profile, are evident.  The longer 
period during which stars show little heating is reminiscent of the 
\cite{quillen01} interpretation of extent observations that
this reflects ``saturation'' in the thin disc's kinematic heating.
The older, kinematically hotter, stars in these simulations have been
suggested to be a signature of the thick disc (e.g., \cite{brook04}).

\begin{figure}
\resizebox{0.5\textwidth}{!}{\includegraphics[angle=-90]{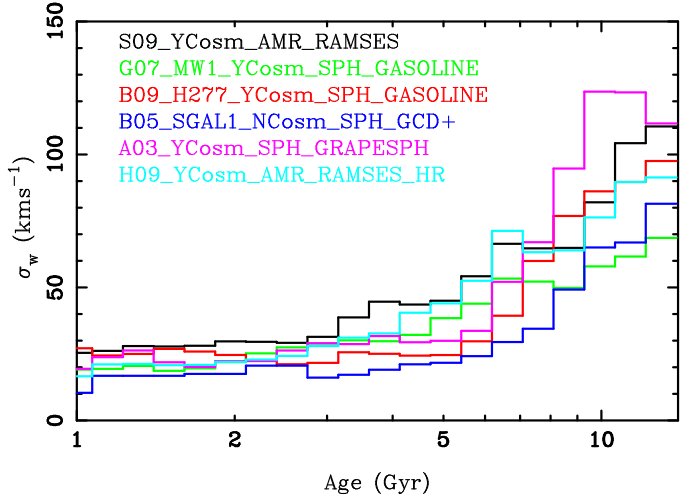}}
\caption{Age-velocity dispersion relation in the vertical direction for 
``local'' disc stars in our suite of simulations.}
\label{fig1}
\end{figure}

Broadly speaking, while there is a continuum of heating ``profiles'' on
display in Fig~1, at one end of the spectrum, several of the 
simulations (e.g., S09, H09, G07(MW1)) show a temporal heating profile which
becomes apparent at younger ages ($\sim$3~Gyr) relative to several at 
the opposite end of the spectrum (e.g., B05(SGAL1), B09(h277)) which
only begin to show evidence of significant heating in older stars
($>$6~Gyr). If we associate these relatively flat periods at 
late times with the thin disc, then none of the simulations have 
thin discs as old as that of the Milky Way, which is considered to be
between 8-10 Gyr old. 

One of the natural consequences of the merging which occurs within the
hierarchical clustering paradigm is a degree of kinematical heating.
As such, we set out to examine 
the merging histories for each of the simulations, to see whether 
they shed light on the characteristic heating profiles and
discrete ``jumps'' seen in the age-velocity dispersion 
plane (Fig~1); these merging histories will be discussed shortly.

One thing which is readily apparent from our analysis is that 
\it all \rm of the simulations show much higher velocity dispersions for the older 
stars in the disc, consistent with the behaviour seen in the Milky Way.  That 
said, there is also a consistent offset, in the sense that all of the simulated discs 
are substantially hotter than that of the Milky Way.  Part of this discrepancy relates to the 
fundamental problem alluded to in \S~1, specifically, that gas cools
efficiently allowing baryons to collapse early during the merging process of 
galaxy formation, resulting in unrealistically large spheroidal components. 
These old spheroidal stars can have a significant impact on the derived 
dispersions, in the sense of ``contaminating'' what one would like to be 
a ``pure'' disc sample. In other words,  rather than measuring the dispersion 
of disc stars, which reflects the observational case, one is instead probing the 
additional impact that the dispersion of the spheroid stars have upon the sample.
This is problematic, at some level, for all of the simulations - 
it is reflected in the very high dispersions seen at large ages in
Fig~1.  In order to make a fair comparison with 
disc stars from the Milky Way, we clean (in a very straightforward
manner) our sample of these spheroidal
contaminants by selecting {\it in-situ} stars, i.e. those that are born in the central 
galaxy. These {\it in-situ} stars are identified as those which form anywhere 
within the central galaxy, while stars that end in the central galaxy at $z=0$ but were 
born within a satellite or substructure are called {\it accreted} . We then derive the 
velocity dispersions of {\it in-situ} stars in our disc defined region, 
7$<$$R_{xy}$$<$9~kpc and $|z|<1$~kpc, at $z=0$.

By selecting {\it in-situ} stars we are examining the heating of disc stars. 
Whether this results in forming the thick disc or merely the old, hot thin disc, 
is left open to interpretation. Of the thick disc formation mechanisms proposed
(see \S~1), the direct accretion of satellites scenario is thus not explicitly 
addressed in this study.  In this  analysis we merely assume that a 
rotationally supported disc forms {\it in-situ} and can be born relatively
hot or cold, and then may be heated by a number of  processes. 
Further, recent simulation results have shown that some 
{\it in-situ} stars will form part of the stellar halo (\cite{zolotov10}, 
and thus may affect our dispersion results. However, these stars 
are in the halo, with too few in our defined disc region to 
affect the dispersion-age plots presented here.

We plot in Fig~2 the age-dispersion relation of these {\it in-situ} 
stars from the simulations 
along with three sets of observational data for the Milky Way disc:
\cite{quillen01}, a combined set from \cite{soubiran05} and \cite{soubiran08},
and \cite{holmberg08}. 
\cite{quillen01} use a sample of 189 nearby F- and G-dwarfs 
from \cite{edvardsson93}; from their resulting
$\sigma$-age relation, they suggest
that the Milky Way disc has been relatively quiescent with little 
heating for stars with ages between 3 and 9~Gyr, with stars older than
that having been subject to an abrupt heating event.
The second set of observational points is taken from \cite{soubiran05} 
and \cite{soubiran08}. We have merged these two catalogues,
in order to include a larger number of old disc stars in the sample. 
We note that this data includes the \cite{reddy03}, \cite{bensby03} 
and \cite{bensby04a} samples, which target the thick disc specifically 
by using a kinematic selection, and this may be the reason that their 
old stars are hotter than in the \cite{holmberg08} samples.
Their analysis is consistent with that of \cite{quillen01},
where the age-velocity relation of the thin disc is characterised 
by the saturation of the vertical dispersion at $\sim$25~km/s at ages
$\sim$4$-$5~Gyr. The final set of observations is that from
\cite{holmberg08}; they present a sample of F- and G-dwarfs
from the Geneva-Copenhagen Survey of the solar neighbourhood 
(\cite{nordstrom04}, GCS) suggestive of a scenario consistent
with continual heating of the local disc throughout its entire lifetime.

\begin{figure}
\resizebox{0.5\textwidth}{!}{\includegraphics[angle=-90]{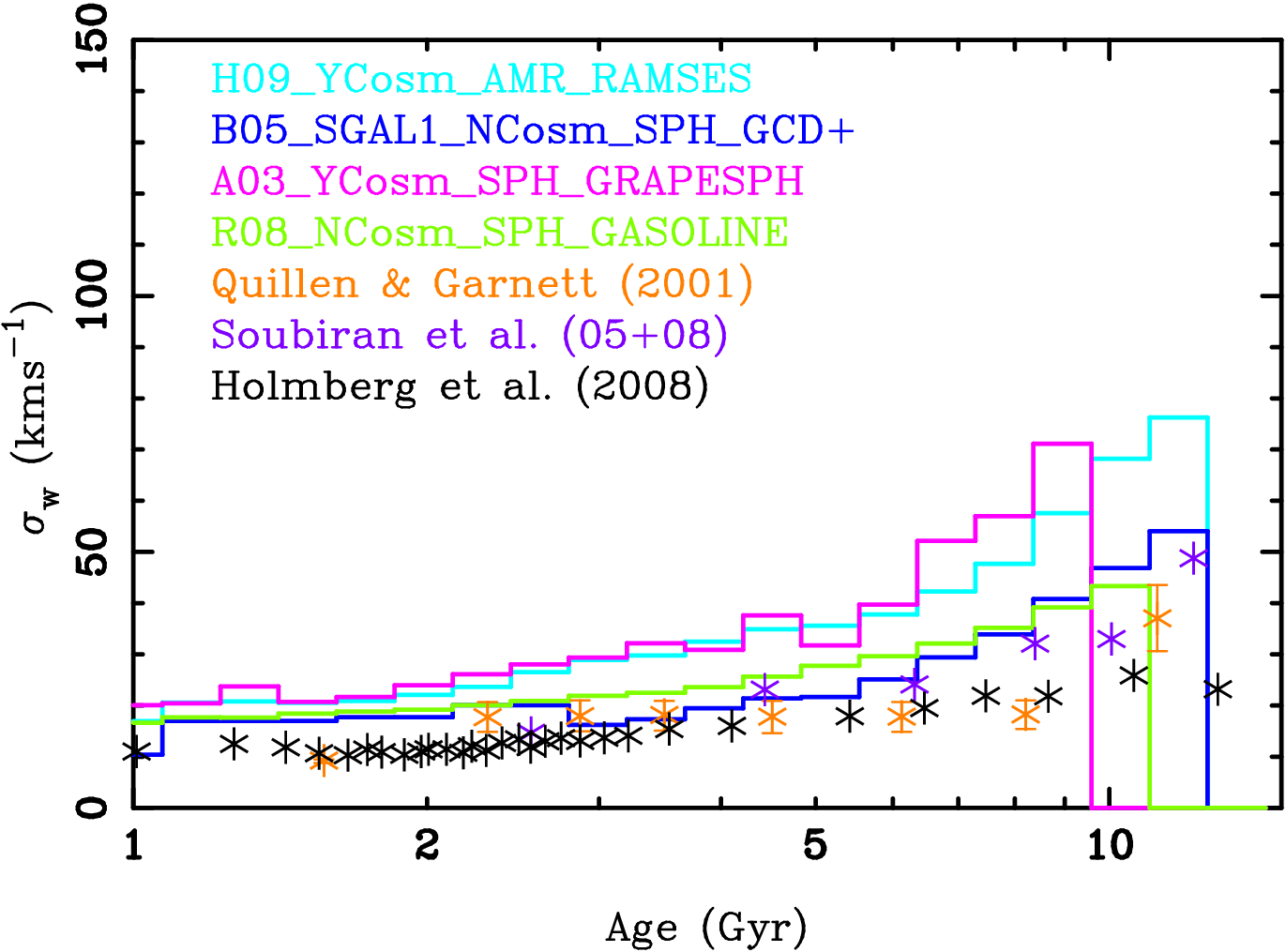}}
\caption{Vertical age-velocity dispersion relation for 
{\it in-situ} disc stars within our suite of simulations,
compared with observations from \citet{quillen01}; \citet{soubiran05}; 
\citet{soubiran08}; and \citet{holmberg08}.}
\label{fig2}
\end{figure}

By only considering {\it in-situ} disc stars, we have 
eliminated a significant fraction of the 
high dispersion old spheroidal components' contaminants; this is 
reflected in the $\sim$20$-$30\% decrease in $\sigma_w$ for stellar
ages in excess at $\sim$7$-$8~Gyr. We have only included four of
the simulations in Fig~2, although the results that follow apply to 
the entire suite.
The overall trend in Fig~2 matches that of Fig~1, in the sense that
a range of heating ``profiles'' are seen, with both continual 
and discrete events being evident.

We also compared the velocity dispersions in Fig~2 in three 
different regions for our highest resolution simulations (so H09, 
R08 and B09(h277), with selected regions being $4<R_{xy}<8$ kpc, 
$6<R_{xy}<9$ kpc, and $7<R_{xy}<9$ kpc). These simulations 
have enough stars to examine much smaller volume cuts. 
We found that the trends in velocity dispersion remain the same, 
independent of radial cut, with quantitative differences with Fig~2 
being insignificant.

It might be argued that the three simulations with the (relatively)
oldest disc component (B09(h277), B05(SGAL1), R08) - 
also, those with relatively flat $\sigma$-age relations for ages 
up to $\sim$6~Gyr - are a somewhat better reflection of the 
relation inferred from the observations.
We will show in \S~3.3 that the merger history of these systems is
a primary driver in the establishment of this relationship, and examine
the role played by numerical effects.

An interesting observation from Fig~2 is that there is a significant 
offset, even when including just {\it in-situ} stars, with all 
simulations compared with observations at any time in the galaxy's 
evolution, in the sense that the stars in the simulations are hotter 
than the observed stars at all times (with the exception of the
semi-cosmological simulation, B05(SGAL1)). This offset is particularly high 
when looking at old stars but is also significant for young stars. 
Several possibilities might be responsible for driving such a 
discrepancy:
(i) numerical heating due to limited force resolution, (ii) the treatment
of heating and 
cooling within the ISM of the simulations, and (iii) the adopted low 
star formation threshold. 

The issue of numerical heating will be addressed at length in \S~4; here,
we simply note that the offset also exists in the 
simulation of \cite{roskar09}, which has a force resolution of 
50~pc, and it also exists in other high resolution isolated disc 
simulations in the literature (e.g., \cite{kazantzidis08}; \cite{kazantzidis09};
\cite{stewart09}). In Fig~2, one can view 
our highest resolution simulation (R08), as well as our lowest (B05(SGAL1));
if numerical heating was the main agent of the observed offset 
between the simulations and observations, one might expect the 
lowest resolution simulation to be (kinematically) the hottest. This is 
not the case though and, in fact, B05(SGAL1) has the lowest resolution and
is the coldest in the sample.

Another important aspect to consider is the effect of secular heating; 
R08 has 
sufficient resolution to account for heating from internal processes such as 
from spiral arms. As the simulation is isolated and 
therefore removed from a cosmological context, the observed heating 
profile in this simulation must be secular due to spiral arms
directly heating stars as well as causing migration (\cite{loebman10}). 
For the R08 simulation, these
internal heating processes \it alone \rm are enough to match the
observations of \cite{holmberg08}.

\subsection{Are Stars Born Hot or Heated Subsequently?}

In this section, we aim to answer several questions that emerged from the 
above kinematical analysis of $z=0$ stars: were the kinematically
hot, old, stars in Figs~1 and 2 born with these high velocity dispersions, 
or were they born ``cold'' and heated subsequently?  If the latter, then
what is the source of this heating?  To answer these, we examine the 
kinematics of disc stars
at the time of their birth for different epochs of a galaxy's 
formation. We do this by selecting disc stars born in the  
``disc'' region, 4$<$R$_{xy}$$<$8~kpc and $z$$<$1~kpc, at the time of their 
birth, using a fairly arbitrary time ``slice'' of 200~Myrs - i.e., we are 
deriving the velocity dispersion of young disc stars in each 
simulation at various epochs. We tested our results with different radial cuts
and age range, and found that our results and 
conclusions are not sensitive to the used values. 
The slightly larger radial slice used in this section allows us to obtain a larger
sample of stars to derive their dispersions. 

Figure~3 shows the derived velocity dispersions for young stars at different
times throughout the respective simulations' evolution. Each of the 
orthogonal components of $\sigma$ are highlighted, although as noted earlier, 
our analysis will concentrate solely upon $\sigma_w$. For clarification,
stars born at early times are situated to the left of each panel in Fig~3, 
while stars born more recently are located towards the right - i.e., the
abscissa now reflects ``cosmic time'' rather than ``stellar age''
(as was employed in Figs~1 and 2).

\begin{figure*}
\resizebox{1\textwidth}{!}{\includegraphics[angle=-0]{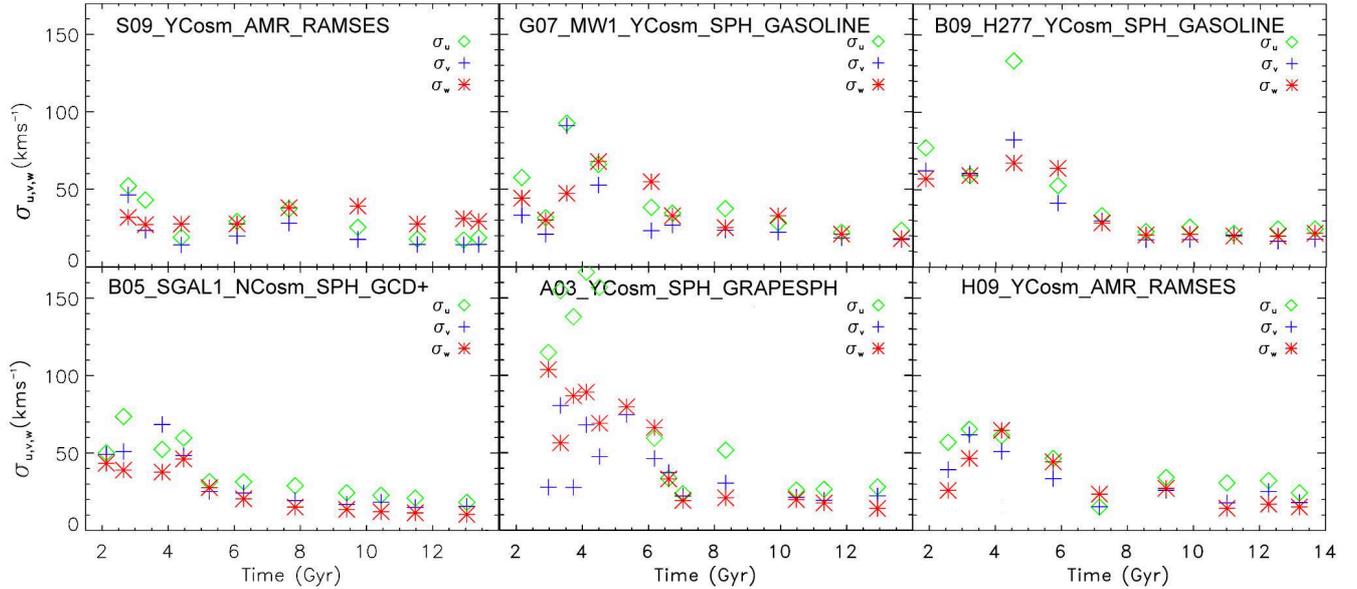}}
\caption{Radial, tangential, and vertical velocity dispersion components
($\sigma_u$, $\sigma_v$, and $\sigma_w$,
respectively) for young stars ($<$200~Myrs) at the time of their birth 
for different epochs, within the defined disc annulus $4<R_{xy}<8$ kpc and 
$|z|<1$ kpc.}
\label{fig3}
\end{figure*}

For the S09 simulation (top left panel of Fig~3), all disc stars, 
independent of time, are born cold with low vertical velocity 
dispersions of
$\sigma_w\approx 30$ kms$^{-1}$, on average. There is a slight increase in 
the dispersion for stars with formation times between $\sim$7 and 10~Gyr,  
where the dispersion increases by $\sim$25\%.  This epoch
corresponds to a period of enhanced minor merger activity, during which
the ISM is heated kinematically relative to the adjoining quiescent
phases.

For the G07(MW1) simulation (top middle panel of Fig~3), stars are born 
on average with vertical velocity dispersions 
between $\sigma_w=20$ and $\sigma_w=30$ 
kms$^{-1}$, except for the period between $\sim$3 and 5~Gyr. 
During this time there are 
several minor mergers with satellites which result in these stars being born 
with velocity dispersions roughly twice that of the adjoining 
phases ($\sigma_w=60$ kms$^{-1}$). It is 
also important to note that these mergers produce a short-lived warp 
at $z=2$ - i.e., at $t=3.2$ Gyr. The stars that we 
detected in the disc during this period were located within this warp
region. Because of 
their potential to dominate over in-plane stars at only a few scalelengths, 
stars in the warp should be treated carefully, particularly in the 
case of studying their kinematics, as they can result in an apparent
increase in the 
velocity dispersion (\cite{roskar10}). These ``warp'' stars are
kinematically ``disturbed'' and born with higher $\sigma_w$.
This is a very similar trend to that
seen in the H09 simulation (bottom right panel of Fig~3) where stars, 
on average, tend to have dispersions between $\sigma_w=20$ and 
$\sigma_w=30$ kms$^{-1}$ at the time of birth, but there is a period 
between $\sim$3 and 5~Gyr, again, 
where this dispersion doubles to about $\sigma_w=60$~kms$^{-1}$. 
As for G07(MW1), this period coincides with an epoch of
enhanced satellite interaction with the main galaxy, although in G07(MW1)
the warp is the primary cause of the high velocity dispersion
during this period and not the minor mergers.  

A distinct trend is noticed for the B09(h277), B05(SGAL1), and A03 
simulations. Stars born at late times (over the past $\sim$6~Gyr) are born
cold, with velocity dispersions between 10 and 20 kms$^{-1}$, while stars
born prior to this are born hot (with vertical velocity dispersion of
$\sim$70~km/s).  It is tempting to
interpret this as the signature of separate thin and thick discs, 
where the thick 
disc is composed of older stars which were born hotter than
the younger, colder, thin disc (\cite{brook04}).

The high velocity dispersion measured at early times in A03, i.e.
from $\sim$2 to 6 Gyr, is due to the numerous merger events that 
this simulations undergoes at early times (see \cite{abadi03a}). 
The feedback mechanism is not particularly effective and therefore the 
satellites that merge with the main galaxy contain a large  
stellar component which affect the high velocity dispersions derived. 
The merger activity is largely over by $\sim$6 Gyr 
and the disc is allowed to settle and form.
One can interpret the low velocity dispersions determined in Fig~3 for
stars from t = 6 Gyr as signature of the formation of such disc.  

Having identified the velocity dispersions of stars \it at birth\rm,
we now wish to determine whether they maintain the self-same dispersion 
as they age - 
i.e., are these stars being heated with time? We do this by 
selecting the same ``young'' stars at a particular time and then
tracing them forward in time, in 
order to quantify the degree of evolution in the velocity dispersion of 
these ensembles of stars.
This is shown in Fig~4, where the subsequent heating of the stars at 
each epoch is represented by the coloured curves.
Because we are interested in stars born in the disc of the galaxy we 
necessarily choose 
epochs after the disc has formed, with the exception of A03 where the disc forms much later
compared to the other simulations. 
For each galaxy this 
time can vary depending upon the time of the last major merger. We 
therefore do not look at stars beyond $z\sim2.5$ because, in general, the 
discs in these galaxies have not yet formed.

\begin{figure*}
\resizebox{1\textwidth}{!}{\includegraphics[angle=-0]{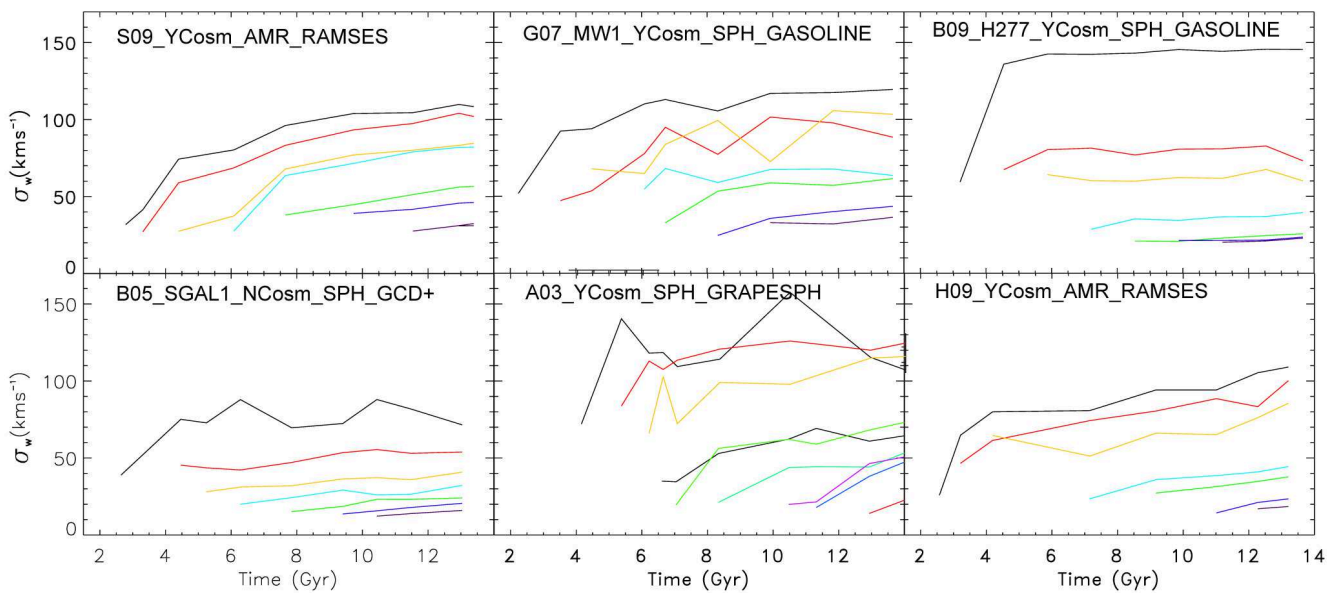}}
\caption{Velocity dispersion perpendicular to the plane ($\sigma_w$)of 
young stars (ages $<$200~Myrs) at various epochs (represented 
by different colours, and different starting times), traced
forward in time to quantify their temporal heating profile.}
\label{fig4}
\end{figure*}

Looking first at the S09 simulation, the stars born at $t=2.5$~Gyr 
have an initial velocity dispersion of $\sigma_w=30$ kms$^{-1}$,
increasing to
$\sigma_w=70$ kms$^{-1}$ over the subsequent $\sim$2~Gyr.
This behaviour is qualitatively repeated for all stars born (and
tracked) in the first $\sim$6~Gyr: i.e.,
stars are born relatively cold but rapidly heat to more than double 
their initial velocity dispersion within $\sim$1~Gyr, before the heating 
begins to ``saturate'', while stars
born over the past $\sim$6~Gyr (while also born ``cold'') 
heat much more gradually.  Indeed, stars born over the last $\sim$4~Gyr
experience essentially no kinematic heating.

G07(MW1) presents a qualitatively similar heating profile to that of S09,
in the sense of (a) older stars experiencing a doubling of their vertical
velocity dispersion in the first few Gyr after birth, before the heating 
saturates, and (b) younger stars experiencing little, if any, 
kinematical heating.  One subtle difference between G07(MW1) and 
S09, though, is that older stars in the former are born relatively hot compared with
their younger counterparts, while in S09, all stars, independent of birth
epoch, are born with essentially the same vertical velocity dispersion.

A03 is similar to G07(MW1) in the sense that older stars are born relatively hot
compared with the younger stars in the disc. The difference in this simulation
compared to both S09 and G07(MW1) is that the younger stars also experience 
heating, where their dispersions double in the first few Gyr. It can be seen 
that even stars born at t = 13 Gyr are being heated with time (see red curve in bottom 
middle panel of Fig~4). 

Both B09(h277) and B05(SGAL1) present somewhat different heating profiles.
With the exception of the first epoch in both (coinciding with the
time of the last major merger in both cases), 
the vertical velocity dispersion is essentially invariant - i.e.,
older stars (which are born hotter than their contemporary 
counterparts, as in G07(MW1)) and younger stars maintain their 
birth velocity dispersion for the lifetime of the simulation.

The next question which needs addressing is this: 
what are the responsible agents driving the assorted heating profiles
seen in Fig~4? What is the quantitative relationship to their 
respective merger histories?  Is the effect of warps playing
an important role?  Are numerical effects plaguing the analysis? We 
address these over the following sub-sections.

\subsection{The Effects of Mergers}

As noted above, our efforts have concentrated upon the
out-of-plane heating within the simulations, due to its 
stronger sensitivity to mergers and interactions.
It is crucial to derive and quantify the merger
histories of our simulations, in order to link the observed heating 
trends with the interactions they have experienced during their evolution.

We have already seen indirect signatures of mergers in the above 
analysis. Stars born with hotter kinematics during early stages of a
galaxy's evolution (Fig~3), as well as the dramatic heating of stars 
born at early times over fairly short timescales (Fig~4), can be 
related to the last major merger (LMM) of each galaxy (see Table 1 for 
the lookback time, $t_{LMM}$, at which the LMM occurred for each simulation).
Major mergers have a total mass ratio of 3:1 or higher.  

In what follows, we examine minor  mergers  of satellites with mass
³ 4\% the mass of the disc at the time of the merger,  back to $z\sim2$. 
Such mergers are able to disturb disc structure (\cite{quinn93}). 
We are, unfortunately,  limited by time resolution due to the available number of outputs for each 
simulation. We are thus not able to trace directly the trajectories of the 
satellites and determine whether they penetrate the disc, 
or the number of close passages which occur prior to the final
coalescence. We restrict our analysis to satellites which have contributed stars to 
 the inner 10 kpc of the central galaxy by $z=0$, indicating that these satellites have 
 interacted with the disc.

The last major merger in the S09 simulation occurs at lookback time of
$t_{LMM}=10.99$~Gyr, which corresponds to a time $t=2.02$~Gyr in Fig~3. 
It has a mass ratio of 3:1 ($M_{vir}$=1.1$\times$10$^{11}$ and 
$M_{sat}$=1.7$\times$10$^{10}$). This major merger heats the disc stars 
significantly as can be seen in the black line in the top left panel in
Fig~3. After the LMM there are several minor baryonic mergers,
the most noteworthy of which occurs between redshifts
$z$=0.8 and $z$=0.7 (a time corresponding to
$\sim$6.3~Gyr in Fig~3). This minor merger has a mass ratio of 8:1
($M_{vir}$=5.7$\times$10$^{11}$ and $M_{sat}$=6$\times$10$^{10}$). 
Stars born during this period are somewhat
hotter kinematically-speaking, relative to those born before and after
(Fig~3). In addition, stars born in the preceding $\sim$3~Gyr to this
merger appear
to have been subject to rapid heating (see the yellow and cyan lines in
upper left panel of Fig~4). Additional (less significant in terms of mass) 
mergers occur at redshift
$z=1.75-1.44$ - i.e., $t=4-4.5$ with a mass ratio of 16:1.
The effects of these mergers are not obvious 
in our plots, although the impact of the former likely plays a role in
the heating seen between times 3 and 4~Gyr in the upper left panel of 
Fig~4. This simulation undergoes minor mergers up to redshift $z=0$.

G07(MW1) undergoes its LMM at lookback time $t_{LMM}=10.89$ Gyr,
which corresponds to time $t=2.5$~Gyr in Fig~3. This major merger has 
a mass ratio of 3:1 ($M_{vir}$=9$\times$10$^{10}$ and 
$M_{sat}$=2.5$\times$10$^{10}$), where the heating caused by this merger
can be seen in the increase in the dispersion for stars during this period
(see black line in top middle panel from Fig~3). It undergoes several 
minor mergers at time $t=3.2$~Gyr with mass ratios of 10:1 and 15:1, the 
effects of which can also been seen in Fig~3 and Fig~4.
Hereafter, it undergoes minor interactions the last one occurring
at $t=7.68$~Gyr. These interactions are apparent  near $t=8$~Gyr in
Figs~3 and 4, although their heating ``impact'' is not particularly
obvious in Fig~3 - i.e., the stars born at $t$=8~Gyr (upper middle 
panel of Fig~3) are not kinematically hotter than those born within
$\sim$2~Gyr of these mergers; similarly, the impact on the heating 
of these stars is not particularly dramatic (upper middle panel of 
Fig~4).  The effect that the LMM (at $t$$\sim$2.5~Gyr) has in
heating the ISM of G07(MW1) is felt over the subsequent $\sim$3~Gyr
(Fig~3), and impacts upon the temporal heating profiles of stars
born during this period as well (Fig~4).  G07(MW1) hosts fairly
significant warps during and after these periods of merger-driven
activity, the impact of which has been noted previously (\S~3.2).

For B09(h277) there is a clear distinction 
between stars born at early times and those formed at later times, 
which can be ascribed to the simulation's merger history.
The LMM in this simulation occurs at a formation time of 
$t_{LMM}=10.41$~Gyr, corresponding to time $t=3.2$~Gyr in Fig~3 and has
a mass ratio of 3:1 ($M_{vir}$=6$\times$10$^{10}$ 
and $M_{sat}$=2$\times$10$^{10}$).    
During the period between $t$=2 and 3.4~Gyr, a significant number of 
both major and minor interactions take place, with a 
final baryonic interaction at $t\sim3.4$~Gyr (with a mass ratio 
of 100:1). This period of merger activity maps directly onto the time
during which the ISM is significantly hotter (upper right panel of
Fig~3). The complete lack of major or minor baryonic mergers subsequent 
to this point is reflected in the absence of detectable temporal heating
in stars born since $t$$\sim$4~Gyr.

We have already discussed the similarity between the heating profiles of
B05(SGAL1) and B09(h277), where there is a clear distinction between stars born at 
early epochs and those born later.  B05(SGAL1) has its LMM at a formation time
of $t_{LMM}=9.74$~Gyr or $t$$\sim$3.3~Gyr in Fig~3 and has a mass ratio of
3:1 ($M_{vir}$=6$\times$10$^{10}$ and $M_{sat}$=2$\times$10$^{10}$.) 
It experiences only one minor 10:1 ($M_{vir}$=6$\times$10$^{10}$
and $M_{sat}$=2$\times$10$^{9}$) baryonic interaction at $t$$\sim$6~Gyr
after the LMM, the signature of which is not readily seen in Figs~3 or 4.

A03 undergoes the last major merger with mass ratio 3:1 at $t$$\sim$6~Gyr 
and it lasts for  $\sim$1~Gyr. The ISM is hotter during this LMM phase, as 
evidenced in the higher $\sigma_w$ at time $t$$\sim$6.5~Gyr in the lower 
middle panel of Fig~3. At times earlier than $t$$\sim$6~Gyr, there are many 
major merger events as can be seen by the large velocity dispersions 
measured for these stars.  It undergoes it last merger event at $z$$\sim$0.74, 
i.e. $t$=7.5~Gyr in Fig~3 and 4  and has completely merged with the disc of the main galaxy
by $z$$\sim$0.48, i.e $t$=9.2~Gyr, with a mass ratio of 45:1
($M_{vir}$=9$\times$10$^{11}$ and $M_{sat}$=2$\times$10$^{10}$).
See \cite{abadi03b} for details of this satellite.
The heating profile of stars formed more recently in
A03 (i.e., those formed within the final 3$-$4~Gyr of the simulation)
differs from those of the other simulations, in sense that even these
recently-formed stars within A03 experience significant heating.
This can be associated to a companion satellite that survives at $z$=0.
It first appears within a radius of 15~kpc at $z$=0.33, i.e., $t$=10.52~Gyr, 
with a mass of $M_{sat}$=5.8$\times$10$^{9}$.

We separate our simulations into two groups, those which have 
interactions at low redshift after the thin disc has formed and those 
that show no major or minor interactions since redshift $z$$\sim$1.
S09 (and H09), 
G07(MW1), and A03, undergo later minor merger interactions and
therefore exhibit more evidence of continual, later, heating.
Conversely,
B09(h277), B05(SGAL1), and R08 (not shown) experience no later minor merger
activity once the disc has formed, and therefore
do not show jumps in the heating over short timescales in their respective
discs at later times, which is associated with such merger 
events in the other simulations.
If we combine this information with what we deduce from 
looking at the age-velocity dispersion plane in Fig~1 and Fig~2, we can 
conclude that in order to obtain a thin disc consistent with 
observations, the simulated galaxy  must experience no interactions 
at late times (at least, since $z$$\sim$1).

\subsection{The Central Concentration of the Satellites}

The effect of heating that accreted satellites have on the disc
is dependent on the mass distribution of the satellite, in the sense
that the accretion of more massive, and more concentrated satellites,
will cause a higher degree of heating (e.g. \cite{velazquez99}).
Simulations produce rotation curves that rise rapidly in the
inner regions with a central peak before dropping off (e.g. \cite{mayer08}). 
However, observations of dwarf galaxies have shown that their
rotation curves rise linearly in the central regions. Presumably, accreted
satellites should have mass distributions which are similar to local galaxies.
The more concentrated satellites in the simulations are related to the
``angular momentum problem'', where the baryons are deficient in angular momentum
and produce overly concentrated stellar bulges.
This challenge for
cold dark matter cosmology is beyond the scope of this paper, but we
note that several mechanisms have been proposed to resolve the
discrepancy between theory and observation (e.g. \cite{navarro96};
\cite{mashchenko08}; \cite{scannapieco09}). \cite{governato10} 
showed that resolution which is high enough to form local star formation 
within an inhomogeneous ISM, will drive large scale supernova
outflows and decrease the central mass concentration, producing
simulated dwarfs which have a mass distribution that matches
observed galaxies. The resolution required to create such dwarfs is not
achieved in any simulation of a Milky Way mass galaxy in our study, or
indeed in the literature.

We looked at the rotation curves of satellites in three of the simulations
discussed in this paper, chosen at redshift $z\sim2$, each with a dynamical
mass of $\sim$10$^{10}$~M$_\odot$ and they showed peaked
rotation curves indicative of an excess of central material. We have
shown that the major source of disc heating in our suite of simulations
is due to the interaction and accretion of satellite galaxies with the
disc. The high central mass concentration of our satellites may be
causing these effects to be exaggerated compared to the effect of real satellites,
particularly if such satellites do indeed have ``cored'' rather than cuspy
central density profiles and no bulge (e.g. \cite{oh10}).
This effect is perhaps the most important caveat to our work; future,
increased, resolution which results in more
realistic dwarfs (akin to those seen in \cite{governato10}) may
reduce the heating rates seen in the current suite of cosmological
simulations.

\section{The Effects of Resolution and Star Formation recipes}

It is important to determine whether numerical heating is influencing 
our results. Two-body heating can have a dramatic effect on the increase 
in the kinetic energy of a kinematically cold rotating stellar disc 
(\cite{mayer04}) and is, therefore, an important factor to take into 
consideration in our study. Such numerical effects are dependent upon 
resolution (\cite{moore96}; \cite{steinmetz97}).
 
In our sample of simulations we have a variety of resolutions. S09, 
G07(MW1), B05(SGAl1), and A03 have relatively low spatial resolution - 
between 400 and 600~pc - while
B09(h277) and H09 have somewhat higher resolution ($\sim$300~pc).
The isolated disc from R08 has a much higher resolution (50~pc).
If the heating we see was dominated by numerical effects, one might
expect a particularly large effect in the lowest resolution 
simulation: B05(SGAL1). In fact, this is the coolest of all the 
simulations studied. Further, we have shown that B09(h277), B05(SGAL1),
and R08 present similar trends in the heating of their disc stars during 
their quiescent period of evolution at low redshift, showing little 
stellar heating, despite having vastly different resolutions.  Of the 
simulations which show significant recent merger activity, H09 is the 
highest resolution, yet it shows heating at low redshift in agreement 
with the lower resolution simulations which have similar merging 
histories (S09, G07(MW1), and A03). These trends appear to indicate that 
numerical heating is not the main driver of the inferred 
heating profiles.

However, the importance of the spectre of numerical heating means that one
must proceed cautiously and examine the issue more quantitatively. In
cosmological simulations, resolution dependence is more 
complicated than the case where isolated discs are used as the initial 
conditions. There is numerical heating due to gravitational softening, 
but on the other hand when we go to higher resolution we resolve more 
substructure, creating more heating. Another problem is that low 
resolution substructures tend to be (artificially) more concentrated 
(\cite{vdBS01}; \cite{barneshernquist96}), meaning that the 
heating effects of their interactions may be exaggerated. We discussed the 
central concentrations of the satellites in these cosmological 
simulations in \S~3.4. 

In order to explore possible numerical heating 
effects, we therefore examined a set of isolated disc galaxies. The 
initial conditions were created as in \cite{kazantzidis08}, to which
the reader is referred for details, and were run 
using {\tt GASOLINE}. The isolated galaxies 
that we re-simulate comprises an exponential stellar disc, a Hernquist 
model bulge (\cite{hernquist90}), and an NFW dark matter profile (\cite{navarro97}).  
The total mass of the galaxy is 10$^{12}$~M$_\odot$, similar 
to that of the Milky Way, with a disc gas fraction of 10\%. To form a 
rotationally-supported disc we impart angular momentum to the 
gas component corresponding to a spin parameter of $\lambda$=0.04. We 
evolve each simulation for 1~Gyr, after allowing 0.2 Gyr for the system 
to relax. The disc has an initial Toomre stability parameter equal to
$Q=2.2$, which means it is stable against any local nonaxisymmetric
instabilities. The main differences between the isolated runs are summarised 
in Table~2. We run three simulations at different resolutions - high 
(ISO\_HR\_LT\_GASOLINE), medium 
(ISO\_MR\_LT\_GASOLINE), and low (ISO\_LR\_LT\_GASOLINE) - and we 
employ the same star formation threshold (0.1~cm$^{-3}$) used in 
the cosmological simulations analysed in this study. The purpose of this 
is to see the effects resolution might have on heating stars in 
simulations. We then run another high resolution simulation, but 
employ a 
much higher star formation threshold (100~cm$^{-3}$: 
ISO\_HR\_HT\_GASOLINE), more comparable to the densities 
associated with star formation, observationally.

\begin{table*}
\begin{minipage}{\linewidth}\centering
\caption{Summary of the Properties for the Isolated Simulations}
\label{Tabel 2}
\begin{tabular}{|c|c|c|c|c|c|c|}
\hline
Simulation Name &  Star Mass       & Gas Mass      & Dark Mass     & $\epsilon^{a}$  & $\epsilon^{b}$  & Star Formation Threshold\\
                & (M$_\odot$)     & (M$_\odot$)  & (M$_\odot$)  &   (kpc)         &   (kpc)         &    $(cm^{-3}$)        \\
\hline
  ISO\_HR\_HT\_GASOLINE & 1.73$\times$10$^{3}$ & 2.56$\times$10$^{4}$ & 1.35$\times$10$^{5}$  & 0.1  & 0.17 & 100 \\
  ISO\_HR\_LT\_GASOLINE & 1.73$\times$10$^{3}$ & 2.56$\times$10$^{4}$ & 1.35$\times$10$^{5}$  & 0.1  & 0.17 & 0.1 \\
  ISO\_MR\_LT\_GASOLINE & 1.38$\times$10$^{4}$ & 2.05$\times$10$^{5}$ & 1.08$\times$10$^{6}$  & 0.2  & 0.34 & 0.1 \\
  ISO\_LR\_LT\_GASOLINE & 1.04$\times$10$^{5}$ & 1.63$\times$10$^{6}$ & 8.68$\times$10$^{6}$  & 0.4  & 0.68 & 0.1  \\
\hline
\end{tabular}
\footnotetext[1]{Dark matter softening length}
\footnotetext[2]{Gas softening length} 
\end{minipage}
\label{isosimproperties}
\end{table*}

In Fig~5, we plot the age-dispersion relation for stars within our set 
of isolated Milky Way disc galaxies. While there is a not surprising 
resolution-dependency in the vertical velocity dispersions of the 
stars at birth (ranging from $\sim$16~km/s, to $\sim$21~km/s, to 
$\sim$29~km/s, for the HR, MR, and LR runs, respectively), there
is little, if any, evidence for any significant heating within any 
of the simulations, 
irrespective of their different resolutions. We will 
return to the differences that resolution has upon the 
stellar velocity dispersions at birth, at the end of this section.

The differences in the velocity dispersions between the isolated 
runs seen in Fig~5 could be due to the effects of feedback in these 
simulations. In order to determine how important this effect may be 
when determining the dispersion in Fig~5, we compare the star 
formation rates (SFR) in all the isolated runs. We find that the 
dispersion is not greatly affected by the feedback in the sense that, 
for example, changes in the SFR (by factors of between 2 and 4) did 
not change the dispersions. Further, comparison between the low and 
high threshold dispersion at times when they had equivalent SFRs, 
we find the same offsets as indicated in Fig~5.

The highest resolution, isolated simulation, like the case of R08, is 
particularly interesting in the context of this section. Both these have 
significantly higher resolution than the cosmological simulations. 
ISO\_HR\_LT\_GASOLINE and R08 are both isolated from a cosmological context, 
and so no heating from satellites occurs, yet these simulations have a 
vertical velocity dispersion ``floor'' of $\sim$15$-$20~kms$^{-1}$, 
similar to the ``floor'' in dispersions seen in the fully cosmological 
simulations during their respective quiescent periods. R08 uses the same star 
formation and supernova feedback physics as G07(MW1) and B09(h277); this
dispersion ``floor'' is tied directly to the implementation 
of ISM physics within
the code.  Such 
physics is difficult to capture in cosmological simulations, as 
it is multi-scale, going from kpc-scale, where most of the gas is 
ionised, to pc-scale, where most of the gas 
is molecular. Cosmological simulations, like the ones analysed here, 
follow the formation of galaxies in a volume of at least several tens of 
Mpc, because aspects of structure formation require that the large scale 
gravitational field is properly modelled. Related to this, our inability 
to resolve locally collapsing high density regions means that we average 
star formation over large columns, using a low density threshold for 
star formation (0.1~cm$^{-3}$ in the simulations analysed here). Yet 
star formation is observed to occur in regions where gas has cooled to 
regions of significantly higher density. The low star formation 
threshold means that, within the simulations, gas may be forming stars 
in regions which remain relatively hot.

Recently, \cite{governato10} showed that with a spatial resolution 
of about $\sim$100~pc, gas could be allowed to collapse to densities more 
representative of the average density observed in star forming
giant molecular clouds.
Using a star formation density threshold of 100~cm$^{-3}$, 
they successfully simulated the first bulgeless disc galaxy (see \cite{governato10} 
and \cite{brook10}).
However, if the density threshold is too high for the resolution
of the simulation, the galaxy comes out too compact. We 
implement these High Threshold (HT) recipes within a high resolution 
isolated simulation (ISO\_HR\_HT\_GASOLINE), and overplot the 
age-velocity dispersion relationship (cyan triangles) in 
Fig~5. Two striking features are immediately apparent: (i) the 
dispersion is much lower than for all the simulations which used a low 
star formation density threshold, even when using the same high 
resolution. The difference is far greater than the difference which was 
caused by resolution.\footnote{see Pilkington et~al. 2010, in prep, 
for a detailed
analysis of the ISM velocity dispersion as a function of star formation
threshold and resolution.} (ii) very little heating occurs, even at 
these very low dispersions. This shows very clearly that numerical 
heating is not affecting our cosmological simulations. Rather, a 
dispersion floor is created by the inability of gas to cool sufficiently 
before forming stars when the star formation density threshold is set 
at a lower level (0.1~cm$^{-3}$).

The resolution-dependent differences in the dispersions of the isolated 
simulations are due to the
differences in the degree to which the gas is able to cool before 
forming stars. This is supported by three pieces of information: (i) no 
heating is apparent at any of the three significantly different 
resolutions, with the age-dispersion relation remaining flat; (ii) stars 
form with lower dispersion at higher resolution; (iii)
even at the very low dispersion levels of the High Threshold simulation 
($\sim$5~kms$^{-1}$), heating was negligible. These 
conclusions are supported by the fact that the temperature of the gas from 
which stars are born increases, as resolution decreases, with 
the average temperature being
7300~K, 6500~K, 5900~K and 400~K, respectively, for the 
LR, MR, HR, and HR\_HT isolated galaxies.

\begin{figure}
\resizebox{0.5\textwidth}{!}{\includegraphics[angle=-0]{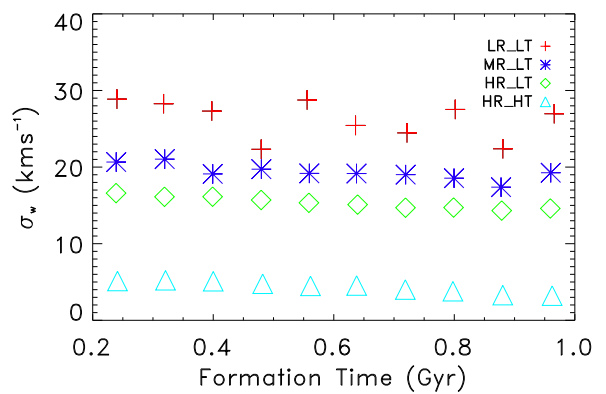}}
\caption{Age-velocity dispersion relation in the vertical direction for 
disc stars ($4 < R_{xy} < 8$ kpc and $|z|<1$ kpc) for four isolated 
Milky Way-scale simulations.}
\label{fig5}
\end{figure}

\section{Summary}

We have analysed the kinematics of disc stars in a suite of 
Milky Way-scale simulations
which were run with different hydrodynamical cosmological codes 
and at different resolutions.
Some were run using the Smooth Particle Hydrodynamics (SPH) 
approach whereas others used the Adaptive Mesh Refinement (AMR) method. 
This is the first paper to compare 
cosmological disc galaxies run with these two different techniques.
No differences in the analysed 
kinematic properties of the simulated galaxies were found to be 
dependent on the approach used for the implementation of gas 
hydrodynamics.

First, we analysed the velocity dispersion of all disc stars as a 
function of age at $z=0$, comparing with analogous observations of
the Milky Way's disc. An overall offset exists, in the sense 
that all the simulated galaxies are hotter than the Milky Way's disc. 
This was shown to be driven in part by resolution and star formation
threshold effects, although the latter is much more efficient 
at reducing the dispersion ``floor'' (Fig~5).  We provide evidence 
that the dominant contributor is the low density threshold
for star formation, 
which has been routinely implemented in simulations of Milky Way-scale
galaxies, although we should point out that we only show this for 
non-cosmological simulations. This low density threshold means that 
stars are formed from unphysically high temperature 
gas, creating a dispersion ``floor''.  Indeed, our lowest resolution 
simulation has the least amount of kinematical heating. 
Future cosmological simulations with sufficient resolution to 
resolve the mean density of giant molecular clouds (akin to the 
\cite{governato10} dwarf galaxy simulations) will be a critical
step forward in this work.

Despite this dispersion ``floor'' in our simulations, 
some interesting heating trends 
are found. Two of our simulations 
(B09(h277) and B05(SGAL1)) are in better agreement with 
interpretations made by \cite{quillen01}, where a saturated 
disc is present for young stars up to $t=6$ Gyr and discrete jumps seen
in the dispersion for older disc stars.
The other simulations (S09, G07(MW1), A03, and R08) seem to be in better 
agreement with the disc undergoing continuous heating, consistent with 
analysis of \cite{holmberg08}, although the rate of heating in the 
simulations remains higher than that observed in nature.

We then proceeded to study the heating of these stars as a function of 
time; starting from the point at which the final disc was stable ($z$$\sim$2),
we derived the dispersion of stars at the time of their birth and how those
coeval ensembles evolved with time.
We found that 
whereas in some simulations stars are born cold in the disc and are 
heated (S09, G07(MW1), A03, and H09), either numerically or due to a 
physical process, in other simulations (B09(h277), B05(SGAL1) and R08) 
the stars maintain essentially the same dispersion as they possessed at birth.
Further, in some 
simulations, stars are born with high dispersions - i.e., they 
are born ``hot''. This 
can be due to interactions (\cite{brook04}) and/or warps 
(\cite{roskar10}). Turbulence in the interstellar medium (ISM) not related to mergers 
could also be a cause of stars being born with large velocity dispersion. 
Recent observations of high redshift discs indicate that internal processes are 
a possible cause of the observed
turbulent ISM (\cite{genzel08}). \cite{bournaud09} compared simulations
formed internally in unstable gas-rich, clumpy discs with simulations of merger
induced disc thickening, and found that thick discs formed internally are a 
better match to observed high redshift discs. Mechanisms such as cold flows 
and supernova feedback are currently being discussed - in addition to mergers- 
as possible causes of the turbulent ISM in high redshift systems
(\cite{burkert09}; \cite{forster10}; \cite{ceverino10}).

Within the favoured cosmological paradigm of hierarchical clustering, 
merging and accretion of satellites onto host galaxies is fundamental.
Our goal has been to examine the effects that these mergers might
have upon the heating of disc stars.
We find a clear relationship when looking at the heating profiles 
between those simulations that have late mergers and those that heat 
significantly. Four simulations (S09, G07(MW1), A03 and H09) have minor 
mergers at low redshift, and we map these interactions onto
the increases seen in the velocity dispersion of their disc stars. 
The other three simulations (B09(h277), B05(SGAL1), and R08), which have 
no interactions over the past $\sim$7~Gyr, show little heating in
the disc with time. We note that R08 is an isolated simulation which has 
no satellites, and hence no interactions, yet has sufficient heating due 
to spiral arms and migration to match observed heating rates of the 
Milky Way. The suite of cosmological simulations do not have the ability 
to resolve these secular effects, nor heating due to molecular clouds. In these 
simulations, it is only in the quiescent period since the last accretion 
events that heating is low enough to match the Milky Way's thin disc. 
None has a thin disc older than $\sim$ 6 Gyr, indicating that it would 
be difficult to gain a thin disc as old as some estimates for the Milky 
Way thin disc within the current cold dark matter paradigm. A caveat of 
our study is the overly concentrated mass distributions of our 
satellites, meaning that resolution of this persistent ``old thin disc" 
problem may come from improved modelling of baryonic physics coupled 
with increased resolution.

\section*{Acknowledgments}
We thank F. Calura for helpful conversations and suggestions during this project.
ELH acknowledges the support of the UK's Science \& Technology Facilities Council
through its PhD studentship programme (ST/F006764/1).
BKG and CBB acknowledge the support of the
UK's Science \& Technology Facilities Council (ST/F002432/1).
The simulations analysed here were made possible by the University of
Central Lancashire's High Performance Computing Facility, the UK's
National Cosmology Supercomputer (COSMOS), NASA's Advanced
Supercomputing Division, TeraGrid, the Arctic Region Supercomputing
Center, and the University of Washington. PSB acknowledges the support of 
a Marie Curie European Reintegration grant within the 6$^{th}$  European
Community Framework Programme. FG acknowledges support from 
the HST GO-1125, NSF AST-0607819 and NASA ATP NNX08AG84G grants.
RT would like to thank the granted access to the HPC resources of CINES and 
CCRT under the allocations 2009-SAP2191 and 2010-GEN2192 made by GENCI.
We thank the DEISA consortium, co-funded through EU FP6 project RI-031513 and 
the FP7 project RI-222919, for support within the DEISA Extreme Computing Initative.

\bibliographystyle{mn2e}

\bibliography{DiskHeating}  

\end{document}